\documentclass[12pt]{article}
\usepackage{graphicx}
\usepackage{lscape}
\usepackage{url}
\usepackage{xspace}

\newcommand\pubnumber{NuPhys2018-Sorel}
\newcommand\pubdate{\today}

\def\ific{Instituto de F\'isica Corpuscular (IFIC), CSIC \& Universitat de Val\`encia\\
  Calle Catedr\'atico Jos\'e Beltr\'an, 2, Paterna, E-46980, SPAIN}
\def\support{\footnote{M.S. is supported by the the Ministerio de Econom\'ia y Competitividad of Spain under grants FIS2014-53371-C04 and FPA2017-82081-ERC, and by the GVA of Spain under grant PROMETEO/2016/120.}}

\def\Title#1{\begin{center} {\Large #1 } \end{center}}
\def\Author#1{\begin{center}{ \sc #1} \end{center}}
\def\Address#1{\begin{center}{ \it #1} \end{center}}

\newcommand\pubblock{\rightline{\begin{tabular}{l} \pubnumber\\
         \pubdate  \end{tabular}}}
\newenvironment{Abstract}{\begin{quotation}  }{\end{quotation}}
\newenvironment{Presented}{\begin{quotation} \begin{center} 
             PRESENTED AT\end{center}\bigskip 
      \begin{center}\begin{large}}{\end{large}\end{center} \end{quotation}}
\def\Acknowledgements{\bigskip  \bigskip \begin{center} \begin{large}
             \bf ACKNOWLEDGEMENTS \end{large}\end{center}}
\def\beq{\begin{equation}}
\def\eeq#1{\label{#1}\end{equation}}
\def\eeqn{\end{equation}}

\def\beqa{\begin{eqnarray}}
\def\eeqa#1{\label{#1}\end{eqnarray}}
\def\eeqan{\end{eqnarray}}

\let\bar=\overbar

\def\Dslash{\not{\hbox{\kern-4pt $D$}}}
\def\dslash{\not{\hbox{\kern-2pt $\del$}}}

\def\msb{{\bar{\ssstyle M \kern -1pt S}}}

\newcommand{\Xe}[1]{\ensuremath{^{#1}\mathrm{Xe}}\xspace}
\newcommand{\Ge}[1]{\ensuremath{^{#1}\mathrm{Ge}}\xspace}
\newcommand{\Bi}[1]{\ensuremath{^{#1}\mathrm{Bi}}\xspace}
\newcommand{\U}[1]{\ensuremath{^{#1}\mathrm{U}}\xspace}
\newcommand{\Kr}[1]{\ensuremath{^{#1}\mathrm{Kr}}\xspace}
\newcommand{\Th}[1]{\ensuremath{^{#1}\mathrm{Th}}\xspace}
\newcommand{\Cs}[1]{\ensuremath{^{#1}\mathrm{Cs}}\xspace}
\newcommand{\Tl}[1]{\ensuremath{^{#1}\mathrm{Tl}}\xspace}

\newcommand{\bbnonu}{\ensuremath{0\nu\beta\beta}\xspace}
\newcommand{\bbtwonu}{\ensuremath{2\nu\beta\beta}\xspace}
\newcommand{\bb}{\ensuremath{\beta\beta}\xspace}
\newcommand{\Qbb}{\ensuremath{Q_{\beta\beta}}\xspace}
\newcommand{\mbb}{\ensuremath{m_{\beta\beta}}\xspace}
\newcommand{\halflife}{\ensuremath{T_{1/2}}\xspace}  
    
\begin{document}
\begin{titlepage}
\pubblock

\vfill
\Title{Xenon TPCs for Double Beta Decay Searches}
\vfill
\Author{ Michel Sorel\support}
\Address{\ific}
\vfill
\begin{Abstract}
Xenon time projection chambers (TPCs) have become a well-established detection technology for neutrinoless double beta decay searches in $^{136}$Xe. I discuss the motivations for this choice. I describe the status and prospects of both liquid and gaseous xenon TPC projects for double beta decay.
\end{Abstract}
\vfill
\begin{Presented}
NuPhys2018, Prospects in Neutrino Physics \\
Cavendish Conference Centre, London, UK, December 19--21, 2018
\end{Presented}
\vfill
\end{titlepage}
\def\thefootnote{\fnsymbol{footnote}}
\setcounter{footnote}{0}
%
%%%%%%%%%%%%%%%%%%%%%%%%%%%%%%%%%%%%%%%%%%%%%%%%%%%%%%%%%%%%%%%%%%%%%%%%%%%

\section{Why Xenon TPCs for Double Beta Decay Searches}

The discovery of neutrino mass and the observed baryon asymmetry in the Universe provide strong motivations to search for lepton number violation. Neutrinoless double beta decay (\bbnonu) is generally believed to be the most promising way to search for lepton number violation and to explore the origin of neutrino mass. Despite a 70-year long history and many null results, the experimental exploration of \bbnonu is experiencing a golden age today. Together with \Ge{76} ones, \Xe{136}-based experiments are providing the most stringent constraints. Xenon can be deployed in large, ton-scale, quantities. In addition, the isotopic fraction of its \bb emitter \Xe{136} can be enriched relatively easily from its 8.9\% natural abundance to about 90\% by centrifugation. Furthermore, \Xe{136} has a high Q-value of $(2457.83\pm 37)$~keV \cite{Redshaw:2007un}, more energetic than many radioactivity-induced backgrounds. Finally, with a half-life of $(2.165\pm 0.061)\times 10^{21}$~yr \cite{Albert:2013gpz}, the two-neutrino double beta decay (\bbtwonu) mode of \Xe{136} has been measured to be particularly slow, hence mitigating the \bbtwonu background contribution to \bbnonu searches.

\begin{figure}[htb]
\centering
\includegraphics[width=0.60\textwidth]{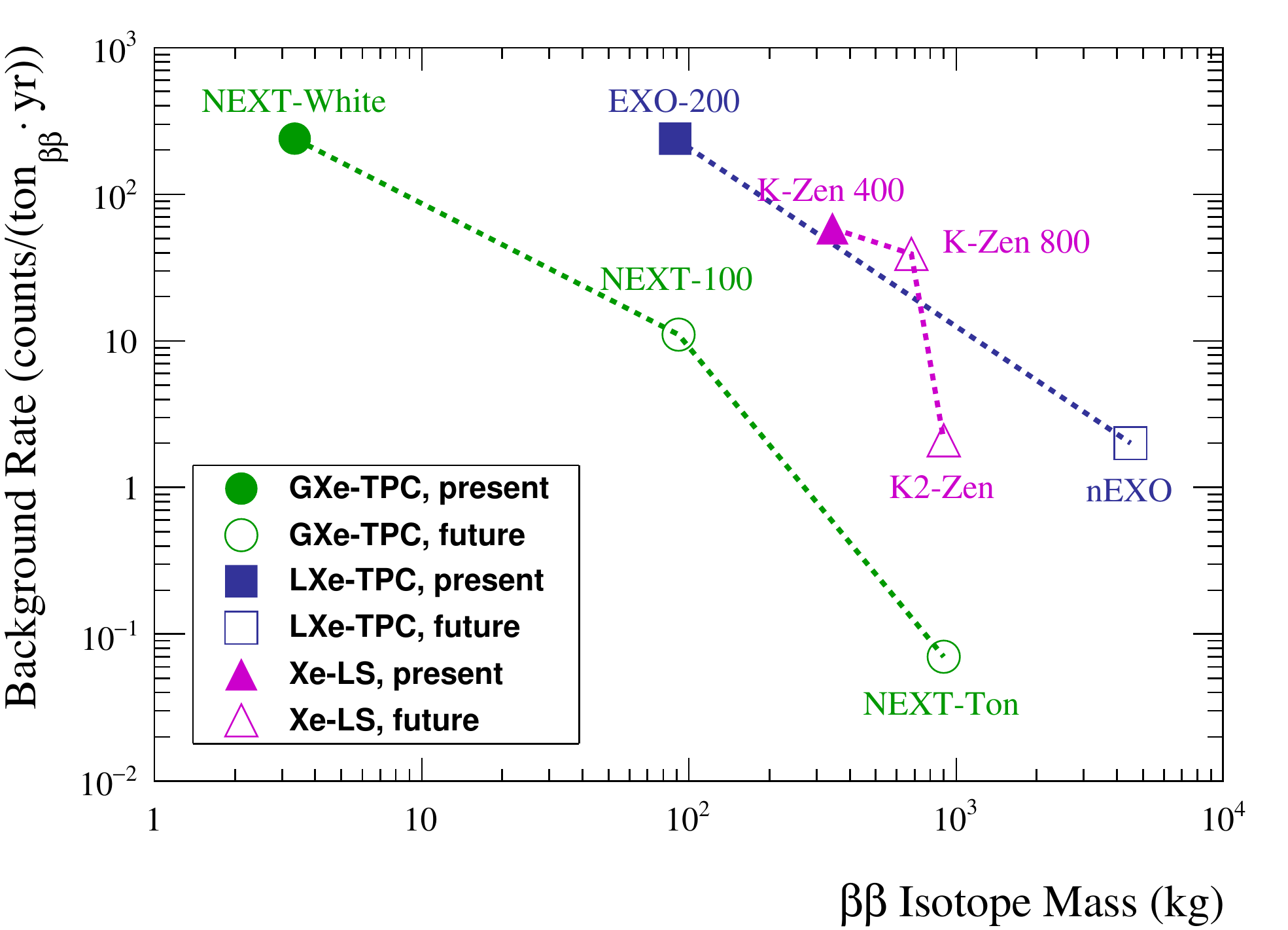}
\caption{Background rate as a function \bb isotope mass, for present and future xenon-based \bbnonu experiments.}
\label{fig:background_vs_mass}
\end{figure}

The three main \Xe{136}-based technologies for \bbnonu are xenon-loaded liquid scintillators, liquid xenon TPCs (LXe-TPCs) and high-pressure xenon gas TPCs (GXe-TPCs). Time projection chambers are easily scalable, particularly LXe-TPCs. In addition, TPCs are expected to provide the lowest project background rates among \Xe{136}-based experiments, particularly GXe-TPCs. This is illustrated in Fig.~\ref{fig:background_vs_mass}, which shows the background rate per unit \bb mass as a function of detector \bb mass, both for current and future \Xe{136}-based detectors. All three \Xe{136}-based technologies shown in Fig.~\ref{fig:background_vs_mass} predict lower background rates per unit mass for increasing \bb masses. This is due to two reasons. First, the external background rate per unit \bb mass is expected to decrease in larger detectors and for the same detector performance, particularly when the detector size exceeds the absorption length of $\simeq$2~MeV gamma-rays. This absorption length is about 8~cm in LXe-TPCs, and about 3~m in GXe-TPCs at 15~bar pressure. Second, background projections shown in Fig.~\ref{fig:background_vs_mass} assume further improvements in radiopurity of materials and/or energy resolution.

LXe-TPC and GXe-TPC detectors have different strengths for \bbnonu searches. LXe-TPCs allow for easier mass scalability and have much better self-shielding against external backgrounds. On the other hand, GXe-TPCs provide a better energy reconstruction and a much more detailed topological signature of the \bbnonu candidates. In addition, and in view of the potential tagging of the barium ion produced in the \bbnonu decay of \Xe{136}, GXe-TPCs have a significantly higher fraction of barium atoms that remain in ionized form, and are hence detectable \cite{Novella:2018ewv}.

%%%%%%%%%%%%%%%%%%%%%%%%%%%%%%%%%%%%%%%%%%%%%%%%%%%%%%%%%%%%%%%%%%%%%%%%%%%

\section{Liquid Xenon TPCs}
\label{sec:lxetpc}

\subsection{EXO-200}

The EXO-200 detector \cite{Auger:2012gs} was a cryogenic TPC with a 110~kg LXe active mass. The TPC consisted of two drift volumes sharing a central cathode. The charge readout was based on induction and collection wire grids, read via electronics located outside the cryostat. The light readout consisted of APDs placed behind the anodes. The detector operated at WIPP during two phases: Phase I (2011-2014) and Phase II (2016-2018). The EXO-200 data-taking ended on November, 2018. EXO-200 accomplished a 1.23\% sigma (2.90\% FWHM) energy resolution at \Qbb, by combining charge and light information \cite{Albert:2017owj}. The detector also used some topological information, to distinguish single-site (signal-like) energy deposits from multi-site (background-like) ones.  

\begin{figure}[htb]
\centering
\includegraphics[trim={0cm 0cm 13.5cm 0cm}, clip, width=0.80\textwidth]{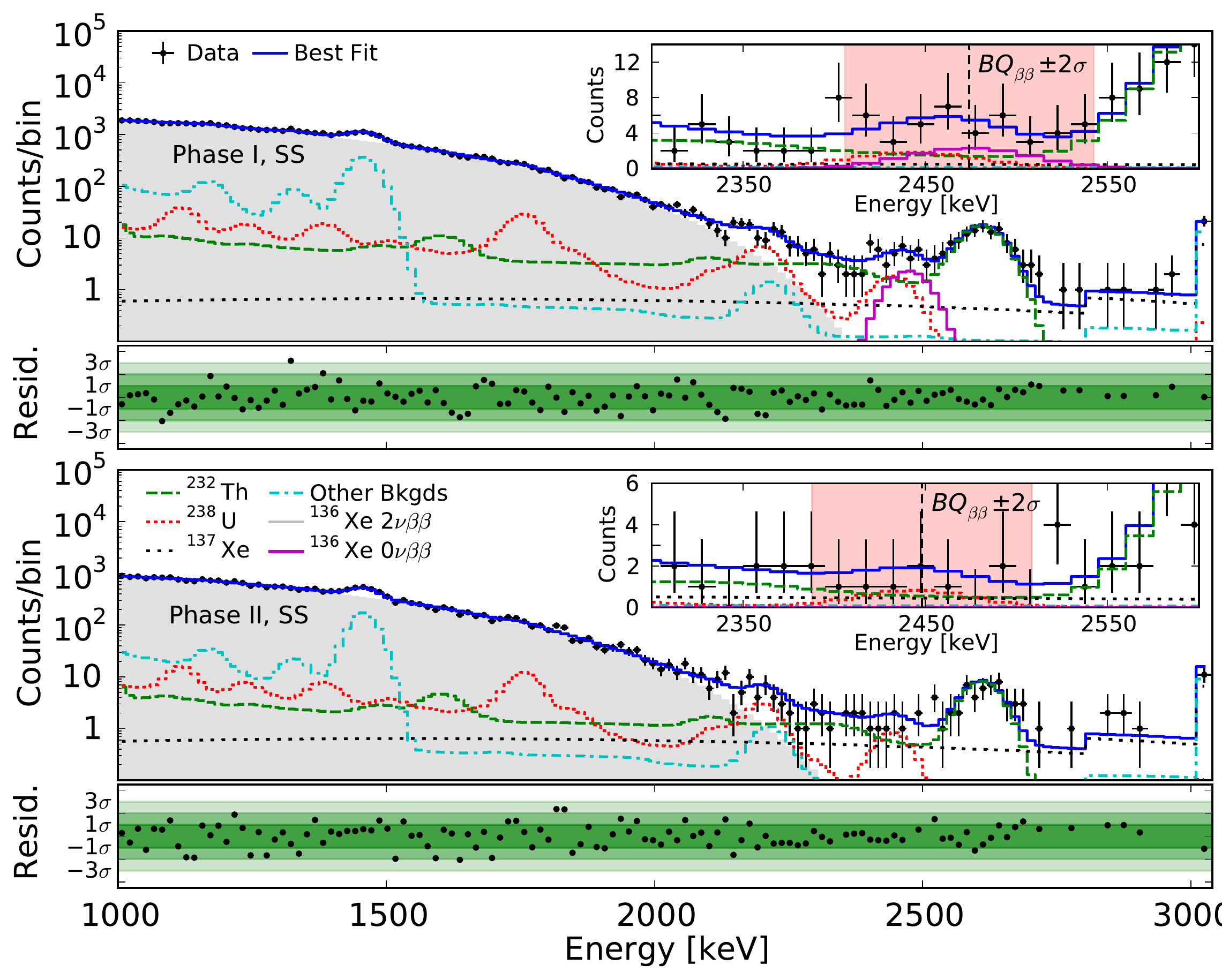}
\caption{Energy spectrum for single-site energy deposits measured in EXO-200 during the experiment's Phase I (top) and Phase II (bottom). The expectations from the EXO-200 fitted background model are overlaid to the measured energy spectra, and residuals (in number of sigmas) between the two are also shown. The insets show a zoom near the energy region of interest for \bbnonu searches. From \cite{Albert:2017owj}.}
\label{fig:exo200_results}
\end{figure}

The detector accumulated a 122~kg$_{\bb}\cdot$yr and 55.6~kg$_{\bb}\cdot$yr exposure in Phase I and II, respectively. The energy spectrum for single-site energy deposits measured in EXO-200 during the two phases of the experiment is shown in Fig.~\ref{fig:exo200_results}. The background rate per unit xenon mass in a $\pm 2\sigma$ wide energy region of interest near \Qbb was measured to be $(0.239\pm 0.030)$~counts/(kg$_{\bb}\cdot$yr). The experiment observed no significant excess above background, setting a \halflife limit of $>1.8\times 10^{25}$~yr at 90\% C.L., to be compared with a median sensitivity of $3.7\times 10^{25}$~yr \cite{Albert:2017owj}. This constraint translates into a Majorana mass limit of $\mbb <147-398$~meV, where the range corresponds to different nuclear matrix element assumptions. 

\begin{table}[t]
  \footnotesize
  \begin{center}
    \begin{tabular}{|c|c|c|c|c|c|c|} \hline
    Experiment & Isotope & Exposure & \halflife Sens. & \halflife Limit & \mbb Limit & Bgr in ROI \\ 
      &  & (kg$_{\bb}\cdot$yr) & ($10^{25}$~yr) & ($10^{25}$~yr) & (eV) & ($10^{-3}$/(kg$_{\bb}\cdot$yr)) \\ \hline
    EXO-200 & $^{136}$Xe & 177.6 & 3.7 & $>1.8$ & $<0.15-0.40$ & 239 \\ \hline
    K-Zen & $^{136}$Xe & 504 & 5.6 & $>10.7$ & $<0.06-0.17$ & 58.5 \\ \hline
    GERDA & $^{76}$Ge & 40.6 & 5.8 & $>8.0$ & $<0.12-0.26$ & 3.4 \\ \hline
    MJD   & $^{76}$Ge & 26.0 & 4.8 & $>2.7$ & $<0.20-0.43$ & 13.5 \\ \hline
    CUORE & $^{130}$Te & 24.0 & 0.7 & $>1.5$ & $<0.11-0.52$ & 388 \\ \hline 
\end{tabular}
\caption{EXO-200 latest \bbnonu results \cite{Albert:2017owj}, compared to other leading published results, from KamLAND-Zen \cite{KamLAND-Zen:2016pfg}, GERDA \cite{Agostini:2018tnm}, MAJORANA DEMONSTRATOR (MJD, \cite{Alvis:2019sil}), and CUORE \cite{Alduino:2017ehq}. Results are compared in terms of \bb isotope, exposure, $T_{1/2}$ sensitivity, $T_{1/2}$ limit, \mbb limit, and background rate per unit $\beta\beta$ isotope mass.}
\label{tab:current_bb0nu_published_results}
  \end{center}
  \normalsize
\end{table}

The EXO-200 result can be compared with the other leading \bbnonu published results in Tab.~\ref{tab:current_bb0nu_published_results}. With 177.6~kg$_{\bb}\cdot$yr, EXO-200 has reached the second-best exposure of current-generation experiments. Despite the relatively high background rate, particularly with respect to \Ge{76}-based detectors, the experiment reached the fourth-best \halflife sensitivity of current-generation experiments.

%%%%%%%%%%%%%%%%%%%%%%%%%%%%%%%%%%%%%%%%%%%%%%%%%%%%%%%%%%%%%%%%%%%%%%%%%%%

\subsection{nEXO}

The nEXO proposed experiment \cite{Kharusi:2018eqi} is meant to be the successor of EXO-200. It is a cryogenic TPC to be filled with about 4~tons of LXe in its active volume. The TPC would consist of a single drift volume along the vertical direction. The charge readout would use charge tiles at the anode, and electronics readout housed in LXe. The light readout would be based on VUV-sensitive SiPMs instrumenting the barrel region of the cylindrical TPC. The nEXO Collaboration submitted a Pre-Conceptual Design Report in 2018 \cite{Kharusi:2018eqi}.

\begin{figure}[htb]
\centering
\includegraphics[height=6.cm]{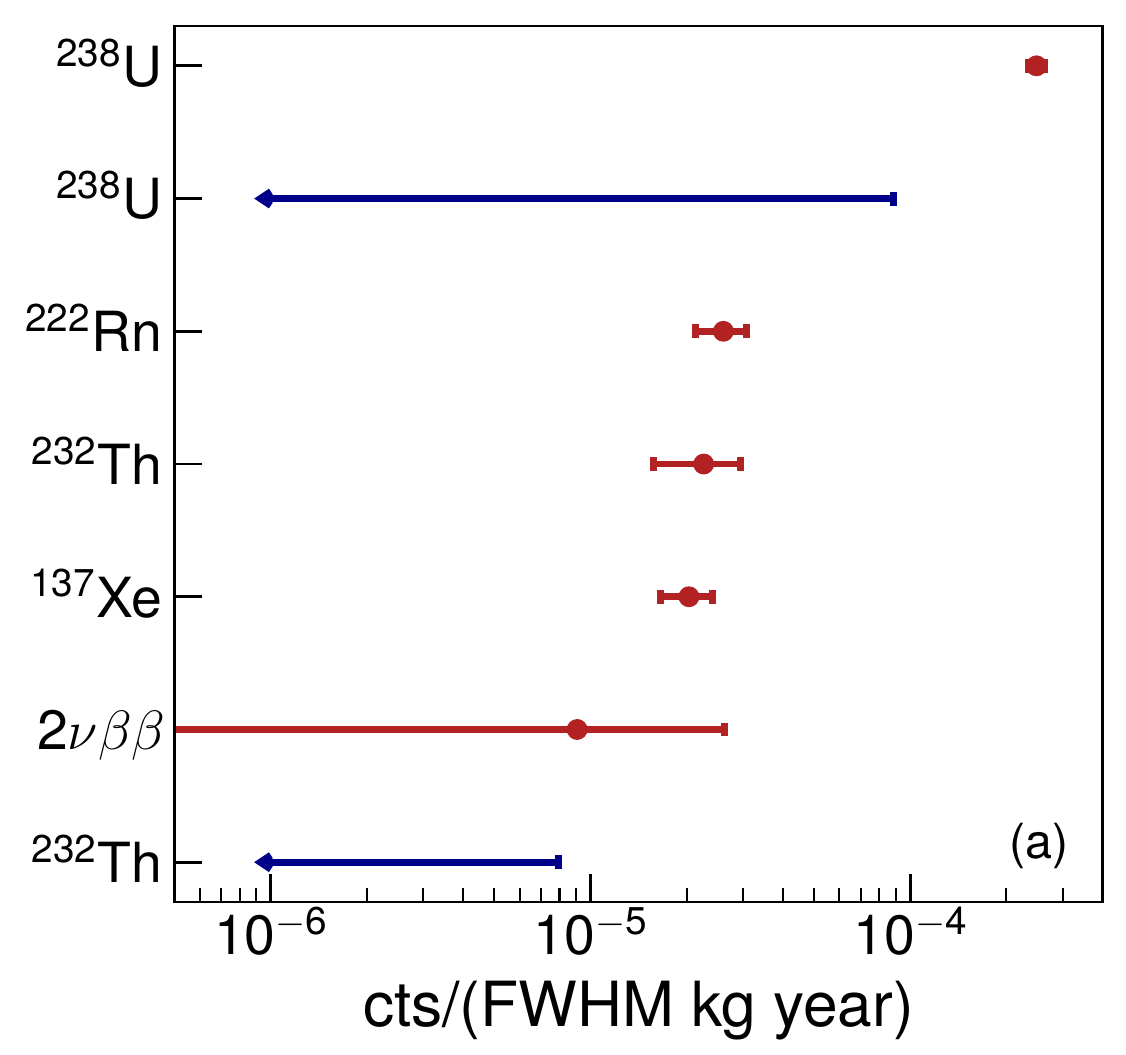} \hfill
\includegraphics[height=6.cm]{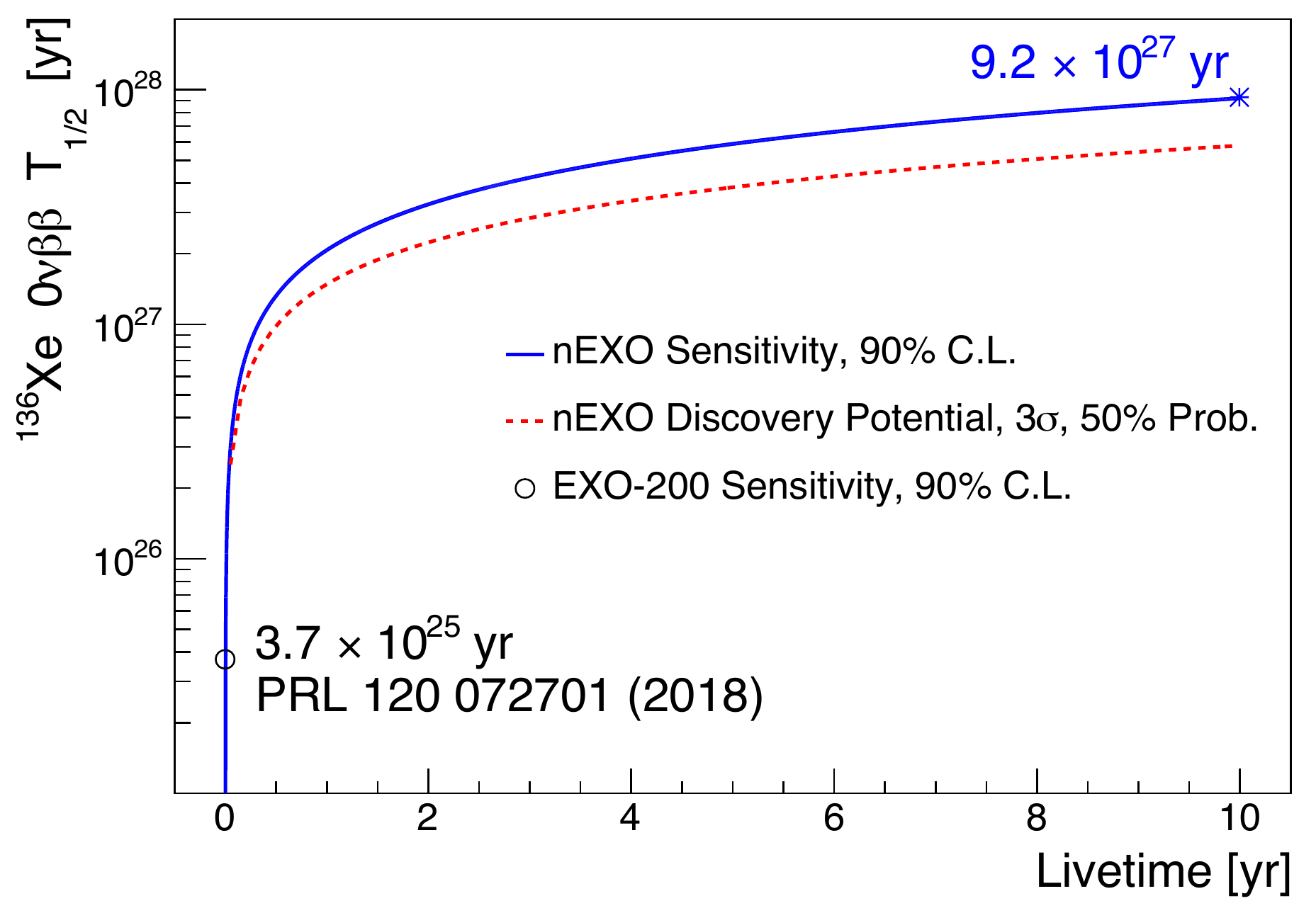}
\caption{Left panel: expected background budget, by nuclide type, in nEXO's inner 2000~kg and for single-site energy deposits with energy within $Q_{\beta\beta}\pm \rm{FWHM}/2$. Right panel: nEXO median sensitivity at 90\% C.L. and 3$\sigma$ discovery potential as a function of the experiment livetime. From \cite{Albert:2017hjq}.}
\label{fig:nexo_sensitivity}
\end{figure}

As shown in the left panel of Fig.~\ref{fig:nexo_sensitivity}, the dominant background in nEXO is expected to be gamma-rays from \Bi{214} produced in the \U{238} radioactive decay chain \cite{Albert:2017hjq}. The most important source of \Bi{214}-induced backgrounds is expected to be the copper vessel. One of the advantages of nEXO is its projected capability to measure very accurately non-LXe backgrounds thanks to the spatial information of the events. The \Xe{136} \bbtwonu background is expected to be sub-dominant even for an energy resolution at the level of EXO-200. The expectation is that nEXO will be capable to improve EXO-200's resolution up to 1\% sigma (2.35\% FWHM) at \Qbb. The \Xe{136} \bbtwonu background contribution will be measured primarily via the energy information of the events. The nEXO project sensitivity shown in the right panel of Fig.~\ref{fig:nexo_sensitivity} assumes 4~tons of LXe in the fiducial volume, and 90\% isotopic enrichment in \Xe{136}. A half-life median sensitivity of $9.2\times 10^{27}$~yr at 90\% C.L. is expected after 10~years of operation \cite{Albert:2017hjq}. This translates into a \mbb sensitivity of 6--18~meV depending on the nuclear matrix element choice, fully covering the \mbb parameter space for the inverted neutrino mass ordering scheme.

%%%%%%%%%%%%%%%%%%%%%%%%%%%%%%%%%%%%%%%%%%%%%%%%%%%%%%%%%%%%%%%%%%%%%%%%%%%

\section{Gaseous Xenon TPCs}
\label{sec:gxetpc}

The gaseous xenon TPC concept for \bbnonu searches is pursued by the NEXT \cite{Alvarez:2012flf}, PandaX-III \cite{Chen:2016qcd} and AXEL \cite{Ban:2017ett} Collaborations. In the following, we focus on NEXT, which is the most developed effort. The NEXT detection concept \cite{Alvarez:2012flf} uses electroluminescence (EL) as a nearly noiseless amplification stage for ionization produced in the xenon gas. The EL (also called secondary, or S2) scintillation light is used for separated energy and tracking measurements. The light is read by two planes of photo-detectors located at opposite ends of the detector cylindrical structure. The energy plane is located behind the transparent cathode, and detects the backward EL light using photomultiplier tubes (PMTs). The tracking plane is located a few mm away from the EL gap, and detects the forward EL light using silicon photomultipliers (SiPMs). The energy plane sensors detect also the primary (or S1) scintillation light produced promptly in the active volume, for event $t_o$ determination. 

During a first phase of the NEXT experiment, 1~kg-scale prototypes at collaborating institutions were operated in 2012--2014 to demonstrate the viability of the detector concept. The 5~kg-scale NEXT-White detector started operations at the Laboratorio Subterr\'aneo de Canfranc (LSC, Spain) in late 2016, and will continue taking data throughout 2019. This is NEXT's first detector to be operated in underground and radio-pure conditions. It has been operated with both \Xe{136}-depleted and \Xe{136}-enriched xenon, in order to measure and understand the background and to measure the \bbtwonu mode of \Xe{136}. The third phase of the experiment is the 100~kg-scale detector NEXT-100, to be commissioned at the LSC in 2020. NEXT-100 and future ton-scale GXe-TPCs will focus on sensitive neutrinoless double beta decay searches.    

\subsection{NEXT-White}

The NEXT-White detector \cite{Monrabal:2018xlr} has a 53~cm maximum drift length and a 40~cm active region diameter, corresponding to about 5~kg of xenon at 15~bar pressure. The xenon is contained in a stainless steel pressure vessel. The tracking plane readout is made of 1,792 SiPMs at 1~cm pitch. The energy plane is made of 12~PMTs providing 30\% coverage of the cathode surface. Radiopure copper of 6~cm thickness is installed inside the pressure vessel, and acts as the innermost layer of shielding. The detector was constructed in 2015--2016, and started operations in late 2016. The detector is continuously calibrated, against spatial and temporal variations of detector response, with 41.5~keV point-like energy depositions from \Kr{83m} decays \cite{Martinez-Lema:2018ibw}. The detector has been operating with electron lifetimes in the 2--6~ms range.  

\begin{figure}[htb]
\centering
\includegraphics[width=0.95\textwidth]{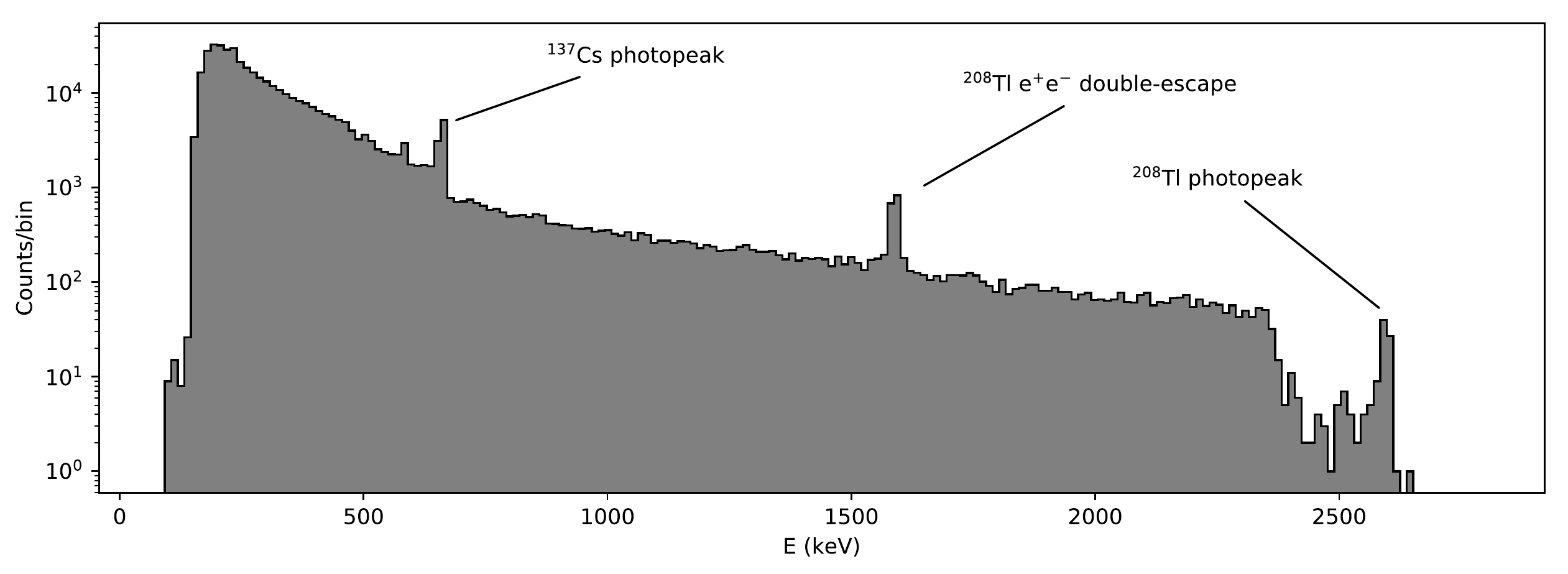}
\caption{Energy distribution of high-energy calibration data in NEXT-White. Updated from \cite{Renner:2018ttw}.}
\label{fig:nextwhite_resolution}
\end{figure}

High-energy calibration has been performed with gamma-ray interactions from \Th{232} and \Cs{137} calibration sources located outside the pressure vessel. The NEXT-White calibrated energy spectrum from high-energy calibration data can be seen in Fig.~\ref{fig:nextwhite_resolution} \cite{Renner:2018ttw}. The \Cs{137} (662~keV) and \Tl{208} (2615~keV) photo-peaks are visible along with their Compton spectra. A peak is also visible at 1593~keV due to e$^+$e$^-$ pair production from the 2615~keV gamma and the escape of both 511~keV gammas produced in the resulting positron annihilation. A study of these peaks demonstrates that the experiment's target resolution of 1\% FWHM at \Qbb has been met in NEXT-White. Calibration events near the \Tl{208} double escape peak at 1593~keV are also used to study how well the topological signature of \bbnonu signal-like events can be reconstructed in GXe-TPCs. Given the low detector density, a \bbnonu appears in a GXe-TPC as an extended single track with Bragg peaks at both track extremes, corresponding to the stopping points of the two electrons emanating from the common \bbnonu vertex. Exploiting the energy distribution information, we can extract how the double-electron (signal-like) efficiency varies as a function of single-electron (background-like) acceptance, for an increasingly tighter double-electron selection. A double-electron efficiency of 75\% for a single-electron acceptance of 22\% is found at 1593~keV from this topological selection in data, with consistent results for simulated data. 

\begin{figure}[htb]
\centering
\includegraphics[width=0.6\textwidth]{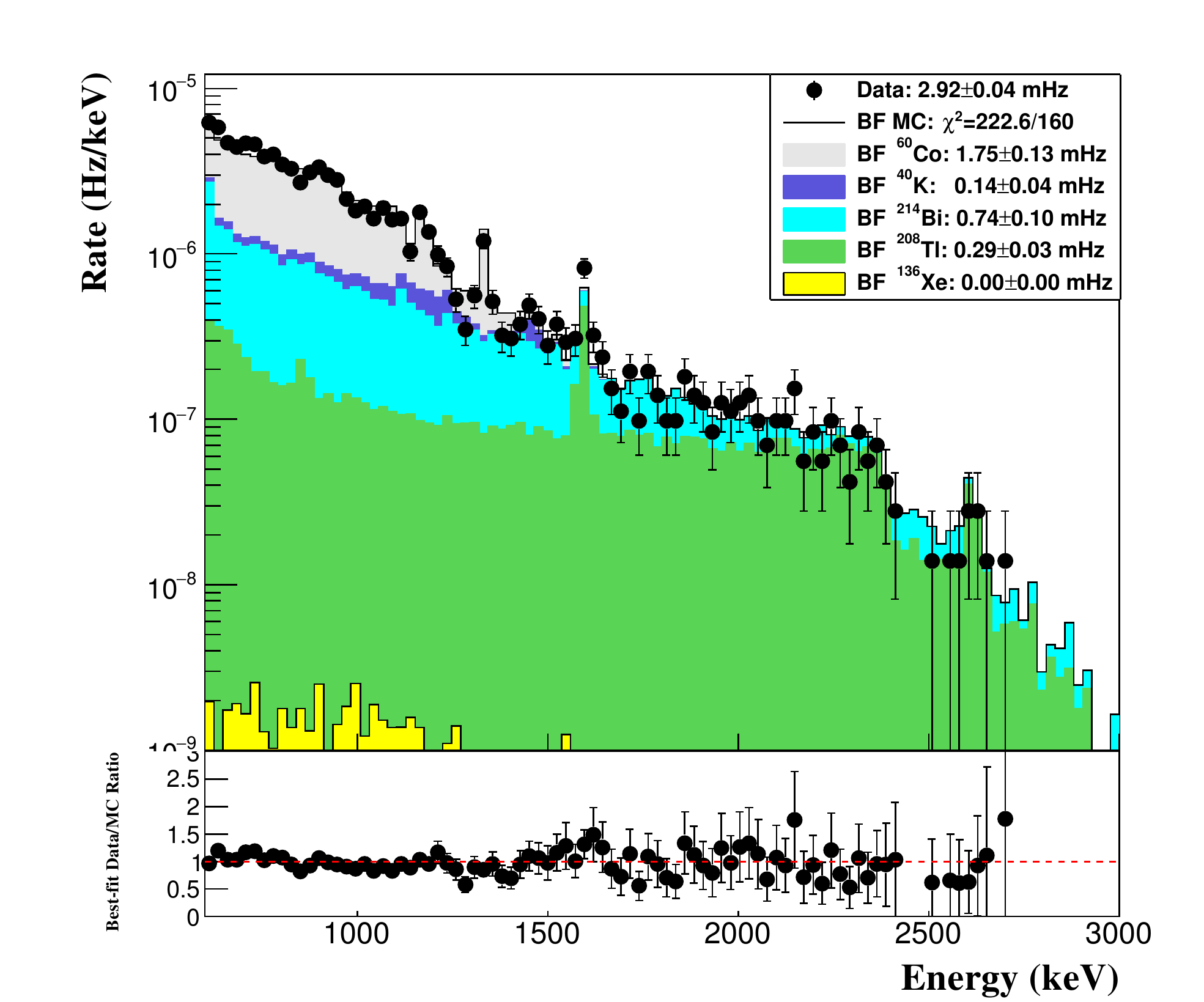}
\caption{Energy spectrum of low-background NEXT-White data with $^{136}$Xe-depleted xenon after fiducial cuts. The data (black dots) are superimposed to the tuned background model expectation (solid histograms). Preliminary results.}
\label{fig:nextwhite_background}
\end{figure}

Low-background operations in NEXT-White started in August 2018, with \Xe{136}-depleted xenon. This period, extending until January 2019, is called Run-IV. During Run-IVa (August--September 2018), a relatively high and variable background rate was observed for $E>600$~keV, due to airborne radon surrounding the vessel. In October 2018, the radon abatement system started to operate, flushing radon-free air to the air volume near the pressure vessel. During this Run-IVb period (October--November 2018), the background rate was greatly stabilized and reduced by a factor of 1.9, on average. In December 2018, additional lead shielding around  the pressure vessel was introduced, to provide better shielding against external backgrounds. As a result, a further reduction in a factor 1.3 in the fiducial background rate was observed during Run-IVc (December 2018 -- January 2019). An accurate background Monte-Carlo (MC) simulation based on extensive radio-purity measurements and a detailed Geant4 description of the detector geometry has also been performed. Figure~\ref{fig:nextwhite_background} shows a comparison of the energy spectrum of fiducial events, prior to any toplogical selection, in Run-IVc data. The data are compared to the MC simulation, broken by isotope type and tuned according to NEXT-White data. Good data/MC agreement is found after tuning, and confirming that no large background source unaccounted for by the radio-purity campaign is present. NEXT-White operations with \Xe{136}-enriched xenon, at a $(90.9\pm 0.4)\%$ enrichment fraction, has started in February 2019. The main objective during this Run-V period is the \Xe{136} \bbtwonu measurement. 

%%%%%%%%%%%%%%%%%%%%%%%%%%%%%%%%%%%%%%%%%%%%%%%%%%%%%%%%%%%%%%%%%%%%%%%%%%%

\subsection{NEXT-100 and Ton-Scale Gas Detector}

The NEXT-100 detector, currently under construction, is a scaled-up version of NEXT-White. It consists of a 130~cm drift length and a 107~cm active region diameter, corresponding to about 100~kg of xenon at 15~bar pressure. The pressure vessel is also made of stainless steel, and the inner copper shield has an average thickness of about 12~cm. About 4,000 SiPMs at 1.5~cm pitch will instrument the tracking plane, and 60~PMTs in the energy plane will provide the same 30\% coverage as in NEXT-White. NEXT-100 commissioning is expected during 2020.  

\begin{figure}[htb]
\centering
\includegraphics[trim={0cm 11.6cm 0cm 1.5cm}, clip, height=5.cm]{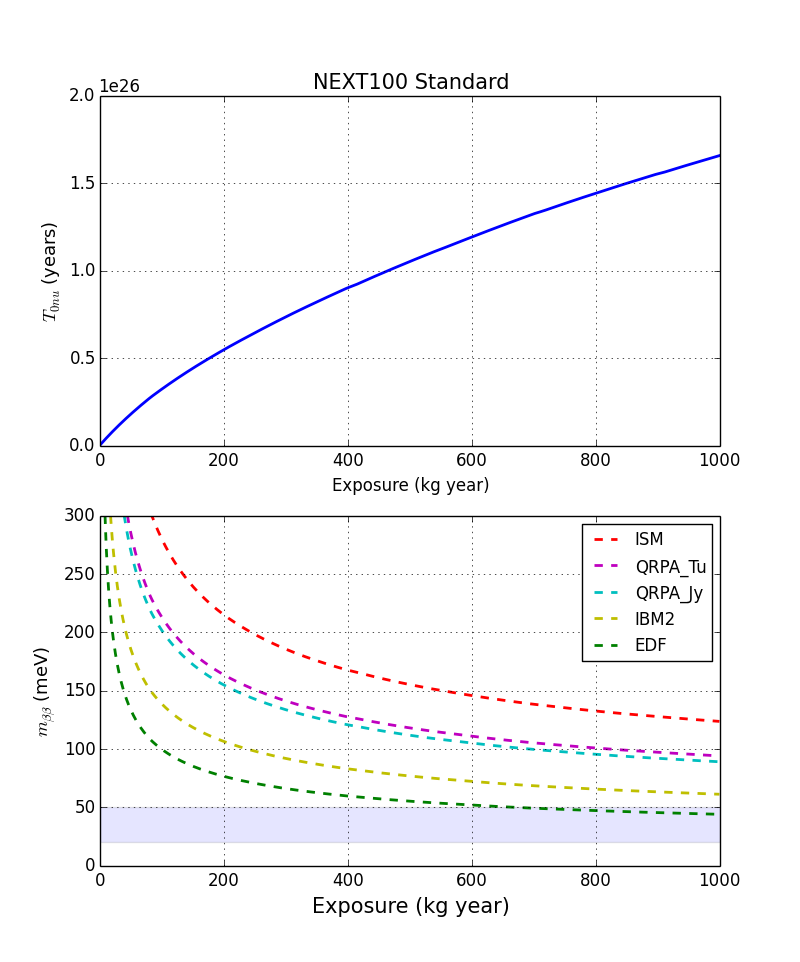} \hfill
\includegraphics[height=5.cm]{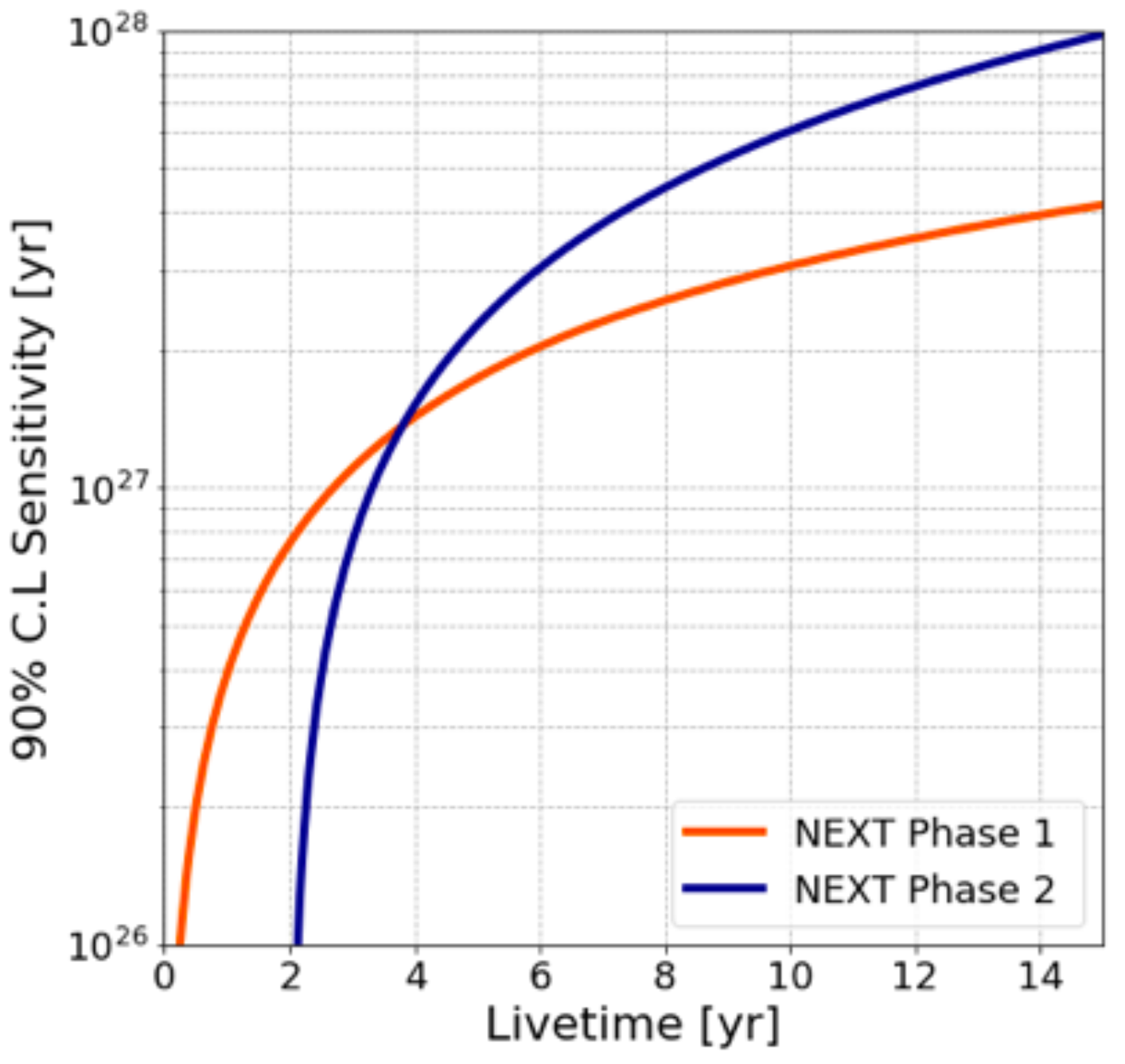}
\caption{Left panel: half-life sensitivity as a function of exposure for NEXT-100. Right panel: preliminary half-life sensitivity versus livetime for a ton-scale gaseous xenon TPC. Two detector phases are shown, with barium tagging capabilities assumed for Phase 2. From \cite{Martin-Albo:2015rhw,munoz_phd}.}
\label{fig:nextfuture}
\end{figure}

The NEXT-100 projected sensitivity to \Xe{136} \bbnonu is shown in the left panel of Fig.~\ref{fig:nextfuture}. A half-life sensitivity of $9\times 10^{25}$~yr is expected after 5~yr of operation and at 90\% C.L., corresponding to a \mbb sensitivity of 70--130~meV \cite{Martin-Albo:2015rhw,munoz_phd}. With one estimated background count per year, NEXT-100 is expected to provide a nearly background-free technology at the 100~kg detector scale. NEXT-100 will also serve as a large-scale demonstrator for a future ton-scale GXe-TPC detector.

The projected reach of a ton-scale GXe-TPC is shown in the right panel of Fig.~\ref{fig:nextfuture}. The figure shows two sensitivity curves. The preliminary design for Phase 1 involves about one ton of low-diffusion xenon gas mixture instead of pure xenon. The vessel and inner copper shielding designs would be extensions of the NEXT-100 design. For the energy readout, PMTs would be replaced by radiopure $5\times 5$~mm$^2$ SiPMs. The tracking plane readout would extend the NEXT-100 design, but would include in-vessel electronics to simplify signal feedthroughs. A water tank would replace the current lead structure for the external shield. A \halflife sensitivity of $1.5\times 10^{27}$~yr, and a \mbb sensitivity of 15--42~meV, could be reached after 5 years and at 90\% C.L. \cite{munoz_phd}.

During a Phase 2, it is envisaged that the detector will be able to effectively tag the barium ion produced in the \Xe{136} \bbnonu decay. Much progress has been made recently toward tagging the daughter Ba$^{++}$ ion via single molecule fluorescence imaging (SMFI) techniques. Single barium ions have been resolved with 2~nm resolution and 13 sigma significance in a aqueous solution already \cite{McDonald:2017izm}. Work is ongoing to demonstrate that the technique can also work in xenon gas. If successful, SMFI-based barium tagging could provide a virtually background-free \bbnonu experiment. It is expected that a ton-scale GXe-TPC could reach a \halflife sensitivity of $3.2\times 10^{27}$~yr after 5~years, translating into a \mbb sensitivity of 10--28~meV.   

%%%%%%%%%%%%%%%%%%%%%%%%%%%%%%%%%%%%%%%%%%%%%%%%%%%%%%%%%%%%%%%%%%%%%%%%%%%

\section{Summary}
\label{sec:summary}

Xenon time projection chambers can meet the two necessary conditions for next-generation neutrinoless double beta decay (\bbnonu) experiments. First, a \bb isotope that can be extrapolated to large, ton-scale, masses. Second, a very low background rate experimental technique, at the level of 1~count/ton$_{\bb}\cdot$yr or less. Both liquid xenon (LXe-TPC) and high-pressure gaseous xenon (GXe-TPC) time projection chambers are being exploited for \bbnonu searches. The advantages of LXe-TPCs are easier mass scalability and more effective self-shielding against external backgrounds. The experimental program is based on the success of the EXO-200 detector, and the nEXO proposed experiment is now in pre-conceptual design report phase. The advantages of GXe-TPCs are a better energy resolution, a more detailed topological signature, and easier tagging of the barium ion produced in the \Xe{136} \bbnonu decay. The GXe-TPC experimental program is being led by the NEXT Collaboration, with the currently operating NEXT-White detector, the NEXT-100 detector under construction, and early plans for a future ton-scale GXe-TPC. 

%%%%%%%%%%%%%%%%%%%%%%%%%%%%%%%%%%%%%%%%%%%%%%%%%%%%%%%%%%%%%%%%%%%%%%%%%%%

\Acknowledgements
I am grateful to G.~Gratta for providing detailed information about EXO-200 and nEXO. I thank my fellow NEXT collaborators, and particularly P.~Ferrario, J.~Mu\~noz, P.~Novella and J.~Renner, for providing several preliminary results presented here.

\end{document}